# Design (not) Lost in Translation: A Case Study of an Intimate-Space Socially Assistive 'Robot' for Emotion Regulation

KATHERINE ISBISTER, University of California Santa Cruz, USA

PETER COTTRELL, University of California Santa Cruz, USA

ALESSIA CECCHET, University of California Santa Cruz, USA

ELLA DAGAN, University of California Santa Cruz, USA

NIKKI THEOFANOPOULOU, Kings College London, UK

FERRAN ALTARRIBA BERTRAN, University of California Santa Cruz, USA

AARON J. HOROWITZ, Sproutel, USA

NICK MEAD, Sproutel, USA

JOEL B. SCHWARTZ, Sproutel, USA

PETR SLOVAK, Kings College London, UK

We present a Research-through-Design case study of the design and development of an intimate-space tangible device perhaps best understood as a socially assistive robot, aimed at scaffolding children's efforts at emotional regulation. This case study covers the initial research device development, as well as knowledge transfer to a product development company towards translating the research into a workable commercial product that could also serve as a robust 'research product' [48] for field trials. Key contributions to the literature include: 1) sharing of lessons learned from the knowledge transfer process that can be useful to others interested in developing robust products (whether commercial or research) that preserve design values, while allowing for large scale deployment and research; 2) articulation of a design space in HCI/HRI (Human Robot Interaction) of *intimate space socially assistive robots*, with the current artifact as a central exemplar, contextualized alongside other related HRI artifacts.

## 1 INTRODUCTION

As HCI practitioners, we are increasingly expected to conduct research that moves beyond testing with a few students in a laboratory setting, into extended field trials with target populations, to demonstrate the validity and efficacy of our designs. This expectation is heightened when we work at the intersection of HCI and adjacent health and wellbeing fields, as the gold standard for demonstrating value is often a randomized controlled trial (RCT), which may require more robust prototypes (and in greater multiples) than we typically generate. Even for initial feasibility evaluations, health and wellbeing HCI prototypes may require a longer-term deployment of devices to see impact, which means that these prototypes must be robust enough to survive ongoing use in field contexts. This article presents a case study of a Research-through-Design [20] process, moving from early research conceptualization through to creation of a robust 'research product' [48] developed in partnership with a product development company, which now makes it possible to conduct larger scale research and longer-term field deployments.

The application space for this project is the scaffolding of children's emotion regulation skills. The team designed and developed a situated and child-led intervention in the form of a responsive 'creature' device with tactile affordances that aid in self-soothing. The knowledge transfer process between our team and the product development company helped us to more clearly articulate the design space that emerged from the research–which we characterize as *intimate space socially assistive robots*.

The article is structured as follows: We begin by describing the research opportunity space and motivations for the research project. Next, we describe the proposed design space for the device, then move on to the design process for crafting the initial research prototype. At this point we begin to introduce the emergent design space of intimate-space socially assistive robots. Next, we discuss workshops with users, and the process of iterating and adapting the initial research prototype into a first research product. This section culminates in a field study that allowed us to validate the efficacy of this first research product. At this point, we discuss issues and concerns with our internally produced research product, and describe the partnership with the product development company that led to producing a more robust, commercially focused device that can also serve as a research product for extended trials of the sort expected by health and wellbeing fields. We discuss lessons learned from the knowledge transfer process, and more thoroughly articulate the intimate space socially assistive robot design space. We conclude with generalizable lessons learned, as well as future work plans.

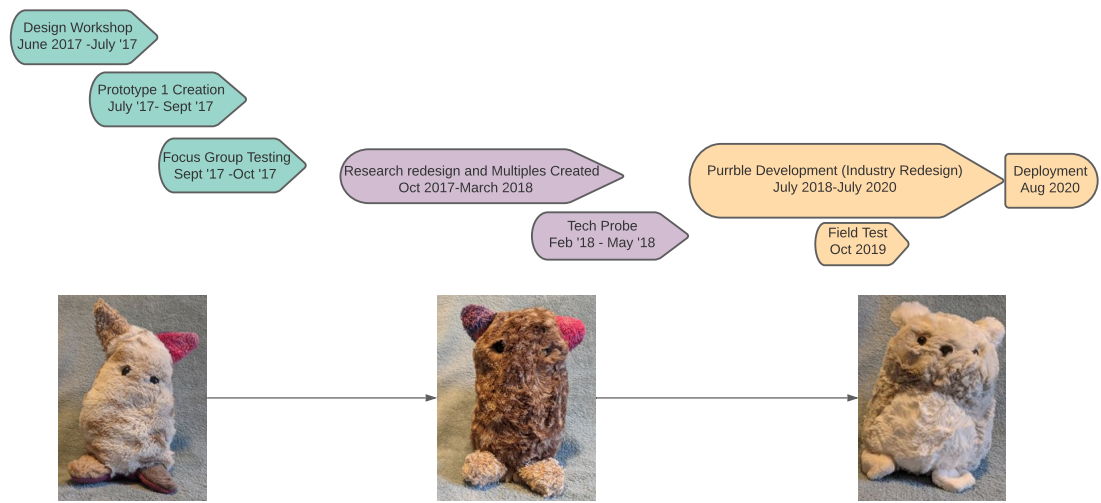

Fig. 1. Timeline illustrating how the research prototypes developed. Prototype 1 was used through the Focus Group testing; Prototype 2 was used throughout the Tech Probe phase; Commercial design was used in a comparative field test and was commercially launched in August 2020.

## 2 RESEARCH OPPORTUNITY SPACE

As noted in the introduction, the application space the team was focused on is the scaffolding of children's emotion regulation skills. Emotion regulation is now well understood as a key protective factor supporting personal well-being (with effects on life outcomes comparable in size to those of IQ or family social status [1, 43]), and is being explored as a likely trans-diagnostic intervention across a range of mental health disorders [7, 47, 53, 57]. Research shows that these effects are wide reaching: if ER is poorly developed, it leads to increased chances of developing mental health disorders [1, 2, 27, 39, 42] as well as societal problems such as criminal behaviour [4], low personal wellbeing [43], and academic under-achievement [15]. In sum, constructive emotion regulation is crucial for maintaining child mental health [14], has links to other important protective life factors such as coping and emotional resilience [8, 64], but can also become one of the strongest risk factors if insufficiently developed [1].

Fortunately, a substantial body of literature in Educational Psychology and Prevention Science shows that emotional-regulation competence is malleable: that is, there are evidence-based interventions that can change children's ability to regulate their emotions in constructive ways (e.g., [15, 66, 68]). While existing programs are relatively successful in targeting children within the 'captive audience' context of schools (see [17, 28, 67] for reviews), a principal challenge remains in extending this support into the day-to-day contexts in which protective competencies are applied, practised, and developed [34]. Moreover, existing intervention research shows that emotion regulation is difficult to develop without detailed in-situ guidance and support [37, 66]; and parenting strategies play a key role in shaping child emotional coping and regulatory skills [16, 24, 36, 41].

To date, however, only limited research in HCI has focused on addressing these difficulties, or to enable new types of interventions that would empower parents and children to further develop protective competencies independently of formal training programs [58, 60]. The Psychology and Prevention Science research outlined here had not taken advantage of the emerging technological opportunities that could enable a shift from what is currently a predominantly in-person program delivery, relying on didactic and role-play based models. What was missing were innovative solutions that would empower children through in-the-moment support and experiential learning – supporting the practice, feedback, and generalisation of skills which appears to be the main challenge facing existing programs [51, 58, 59, 68].

## 3 PROPOSED DESIGN DIRECTION

The team drew upon extensive prior interviews with parents and children about their emotion regulation strategies and family communication about emotion regulation [60] to develop an overall approach for the kind of intervention that we wanted to scaffold with the device: 1) the device would be a 'situated intervention' allowing children to practice regulation in the moment in everyday life, and 2) the intervention would be child-led rather than parent driven.

To meet these criteria, we leveraged a common everyday practice for helping children manage emotion and attention in the classroom–the use of fidget objects. There is a cottage industry that provides specially designed fidgets for this purpose (e.g stress balls and the like–see [56]). With this research-through-design project, we asked: What if we were to take advantage of this natural tendency (fidgeting with objects to aid in self-regulation) to create an intentionally-shaped opportunity for scaffolding development of emotion management skills in children? There has been preliminary work establishing links between particular fidget characteristics and adult and child patterns of use toward self-soothing [12, 35], for example, soft surfaces with give that feel good to squeeze or stroke when angry or anxious. There has also been preliminary work exploring the potential of soft-bodied smart fidget objects to provide appropriate affordances while capturing touch traces, toward the creation of interactive fidget devices [9, 23]. However, no one had yet systematically developed a soft-bodied smart fidget device explicitly for this application space. Thus, the research team set out to craft a device specifically for children (focusing on elementary school age as a crucial target for preventative interventions, cf., [60] for details), which offered the child an opportunity to engage in specific fidgeting interactions, toward scaffolding their own self-soothing and emotional regulation practice.

We were struck by results from prior work [12] in which children were asked to imagine and explain their own ideal fidget devices. Many of the children produced sketches of creature-like fidgets, with eyes, limbs, and furry surfaces. We realized that children were very familiar with hands-on play with toys that had animal-like properties, and that this helped to set a familiar social frame for interacting with the toys. At this early moment in the design process, we decided to leverage the social expectations for interaction that would go along with framing the fidget device as a creature. This led us to specify a role for the device, and to develop it as a socially assistive 'robot' tangible. The role we selected was a small, vulnerable creature (not unlike the child) but also, with the potential to develop coping strategies (like the child). The creature would have an ambiguous identity and backstory that could allow projection by the child onto the robot. This could facilitate the child to rehearse coping strategies under the guise of caregiving. In this we drew upon prior research showing the therapeutic benefits of interaction with creatures that are vulnerable and smaller than oneself (such as pets–see [55] and [11]), and the benefits of role-play as a path toward self-efficacy and empowerment in situations a person perceives as difficult or scary [65]. We also drew upon research exploring the benefits of strategically deployed vulnerability in technology design [13], and on the use of ambiguity in design to allow for user projection and meaning making [22].

Thus the intended core logic of interaction with the device would be that the child would be able to sense that the creature was anxious, and would be able to use (self-soothing) fidget behaviors to calm it, leading the child to engage in emotion regulation strategies well known to psychology researchers of *attention redeployment* (from their own feelings/troubles to those of the creature) and *response modulation* (engaging in self-soothing fidget behaviors under the guise of caring for the creature [60]) – see [61] for a detailed description of the underlying intervention theory of change.

Our intent was to create an interactive device (now framed as a socially assistive robot) that used a particular role (vulnerable, small creature) to evoke self-soothing behaviors in the form of care-taking the creature. Guiding design concepts we brought to this process were allowing for **ambiguity**[22] in the design, to allow children to project their own concerns and stories onto the device; and evoking **vulnerability**[13, 31] to encourage care-taking and long-term engagement. We also intended to design the creature so that it could be seen as having the **potential to develop coping skills**. The device also needed to afford appropriately **self-soothing tactile interactions**.

## 4 DESIGN PROCESS

The design process involved interrelated decisions about the exterior and interior of the device that would: 1) enable users to **project** an appropriate persona and invite a particular social relationship, 2) **invite** appropriate self-soothing tactile interactions, and 3) **elicit** (through appropriate feedback) ongoing tactile interactions and social interpretations from the child that scaffolded their emotion regulation. Here we break down iterations of each of these design factors:

### 4.1 Projecting a persona/relationship

Early in the design process, we settled on a desired relationship between the child and the creature. The child should feel that they could be a caretaker for the creature, supporting it in calming down. But at the same time, the creature should not seem as helpless as an infant. Instead, the creatures should seem as if it could learn to master its emotions, and become empowered. So the creature should seem vulnerable, but also capable [33]. In terms of what sort of creature it would be, there was interest in providing soft, fuzzy fidgetable surfaces–so we never envisioned the creature as humanoid. There was concern about making the creature any particular sort of animal, because there could be strong associations with real creatures that children might map onto the device (including particular learned likes and dislikes and expectations of complex behaviors). We aimed for an ambiguous creature that had familiar affordances (face, ears, limbs, maybe a tail) without evoking a particular species. We knew from the beginning that we did not want to give the creature a mouth, as we did not intend to engage the child in conversation, and we did not want to settle a clear expression on the creature's face. Rather we wanted its emotional state to be read from its haptic feedback alone.

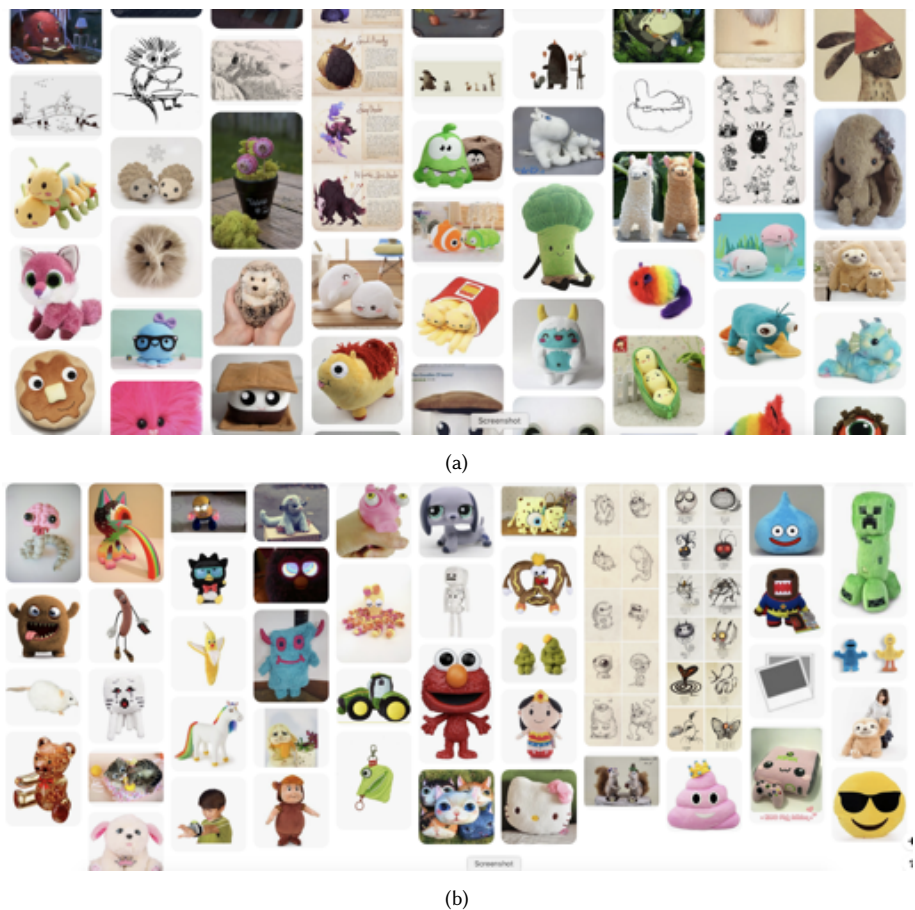

(a)

(b)

Fig. 2. Mood board images used in design process for desired (2a) and undesired (2b) qualities.

To help the team ideate the creature's appearance, we created positive and negative mood boards (figure 2) of existing creature images and toys that had features like those we intended to create. The positive mood board showed cute, but also not too baby-ish creatures, with an element of wilderness. The negative mood board included overly baby-ish or emoji-style characters. We moved from these to sketches of various creature ideas (figure 3). Some of these were hybrids of familiar creatures which we did mock up (figure 4), but we abandoned this approach in favor of more completely ambiguous creatures. The final prototypes of the anxious creature from this first phase (figure 5) were used in the focus groups presented below. The mottled fur suggests that the creatures have been living on their own for a while, and have some kind of independent existence. We settled on a very simple face without clear expression, and we included ears, feet, and a tail to provide a range of fidgeting surfaces with different affordances. The creature posture was a neutral one, onto which a child could project a variety of emotional states (e.g. not fully relaxed or prone, not standing upright, leaning neither forward nor backward in approach or avoidance). In these initial prototypes, we experimented with providing the creature with a bit of clothing (left, a traveling cloak) to underscore its autonomy and independent life and adventures, and even with a smaller companion (right, riding atop the head) to mirror the relationship between the child and the creature.

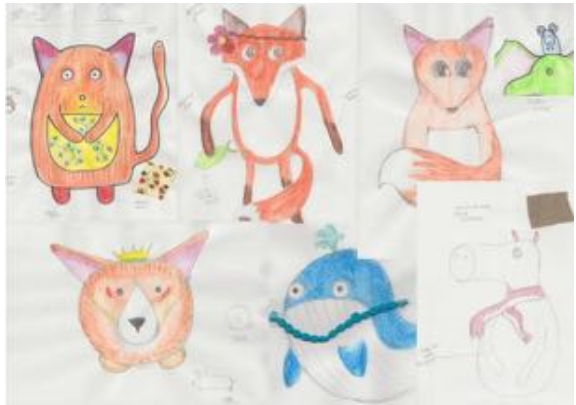

Fig. 3. Sketches of creature ideas.

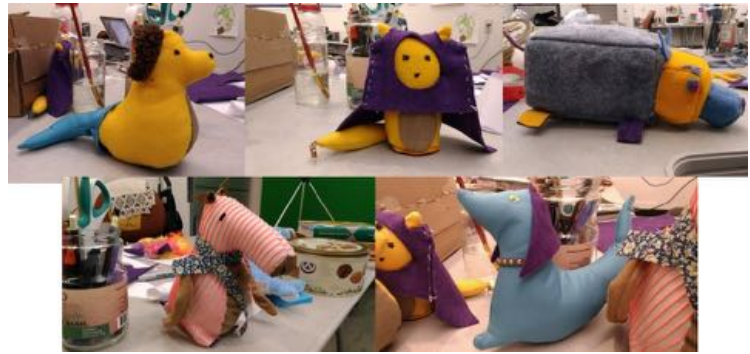

Fig. 4. 3d mockups of creature ideas.

## 4.2 Inviting tactile affordances

Here we drew upon prior research examining a variety of soft-bodied fidget affordances [9], as well as work that collected self-reported fidget behaviors from children and their caregivers, exploring links to emotional and cognitive regulation [12]. The latter work emphasized the importance of providing squeezable surfaces, as this was a frequent fidget behavior for children in times of stress. So, the body of both prototypes included an internal pressure ball sensor, embedded in a soft batting that invited gentle squeezes/hugs. We constructed the creature ears from a copper mesh that could be manipulated and slightly reformed, that also had pressure sensors. We imagined that children might manipulate the ears into shapes that they thought mirrored the creature's emotional state as part of their play. Because children also really enjoyed clickable fidget surfaces in the prior research [12], we put clickable buttons into the creature feet. We also gave the creatures a flexible tail which could be squeezed and gently manipulated, that had some stiffness so as to aid in the creature's balance. As with the ears, we thought children might reposition the creature's tail to help indicate how it was feeling. Each of these affordances provided sensor data to a central processor. To keep the creature small enough for a child to hold it, and light enough for them to easily manipulate it, we chose to use Adafruit Feather development boards. We selected the FeatherM0 for the microcontroller board, and added FeatherWing attachment boards with a real-time clock and micro-SD card so that we could log time stamped data in the event of multi-day, in-home trials.

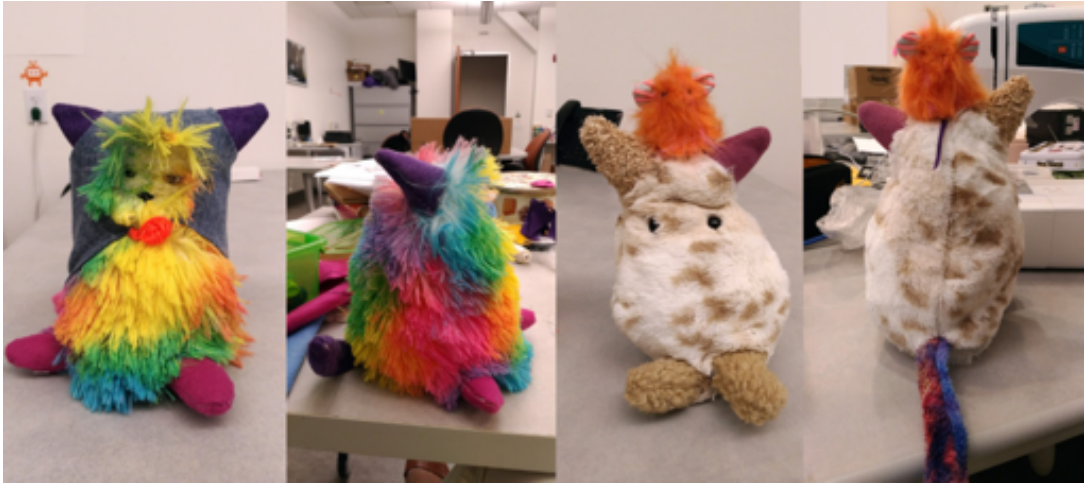

Fig. 5. First complete prototypes of anxious creature concepts.

### 4.3 Eliciting ongoing interactions

The key to scaffolding children's emotional regulation, we postulated, would be providing an ongoing interaction between the child and the creature that evoked self-soothing fidget patterns from the child as a way to engage the creature. We crafted a backstory in which the creature arrived one day on the child's doorstep, and was easily scared and nervous from its (unknown and mysterious) past experiences. The creature would 'wake up' anxious, and would need to be soothed by the child using tactile interactions which also happened to be soothing for the child as well. These gentle tactile actions would eventually put the creature into a happy, relaxed state.

In developing the interactivity, it was helpful to make use of the MDA (mechanics, dynamics, aesthetics) framework from game design [30, 32], conceiving of the interaction between the child and the creature as a kind of game loop. The child is motivated to calm the creature, and takes actions to soothe it. The creature's feedback helps to guide the child's actions toward self-soothing behaviors. So, the aesthetic in this case is a mutually soothing interaction that builds a sense of competency in the child in their caregiving of the creature, which also leads the child to feel calmer. The mechanics of the interaction are purely touch-based (activating various sensors through manipulation of the creature). The resulting dynamics that we crafted were haptic-motor-based state changes–the creature 'woke up' with a rapid 'heart beat'. Soothing touches would gradually slow the heart beat, which would eventually change to a gentle 'purr'. In this early prototype, it was necessary to press one of the foot buttons to 'wake' the creature and begin the interaction (so as to avoid draining the battery with extended 'listening' for touches when the device was not being used). If the creature was only 'woken up' without further interaction it was on a timer that stepped the heart beat down until the motor was off or 'asleep'. This cycle would happen sooner if the creature was interacted with in a soothing manner.

The final creature design as a result of this first stage, which we used in initial focus groups with children, is shown in figure 6.

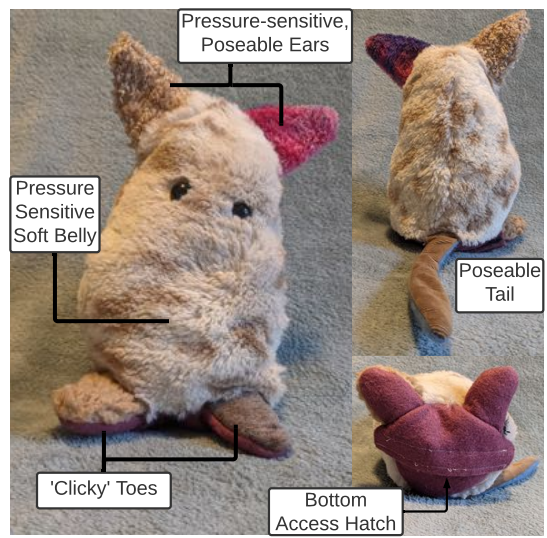

Fig. 6. Anxious creature research prototype used in workshops with children.

## 5 BEGINNING TO ARTICULATE A DESIGN SPACE

In constructing our first prototype, we found that we were evolving a notion of an 'intimate space' set of interactions, in the sense of Edward Hall's study of proxemics [29]. Hall observed that interpersonal interactions took place at varied distances, and he characterized these as public, social, personal, and intimate space (see figure 7). We decided all interaction with the anxious creature should be focused on the intimate space–bringing the creature very close, handling it, even hugging it. Developing this notion helped us to get clear on what the creature would not do–it would not engage the child in conversation, for example, which takes place in the personal and social space zones. This meant we did not need to focus on developing facial expressions or gestures for the creature. Rather we could focus on what could be best sensed in the intimate zone–touch and vibration. There is precedence for using proxemics as a framework in both play design [45] and in evaluating human robot interaction [46, 49]. While game designers have at times focused on intimate-space interaction [45], we did not find human robot interaction characterized in this way in any prior work.

Instead, proxemics focused work concerning robots has so far looked more at human behavior at the public, social, and personal space distances (e.g. [46, 62]).

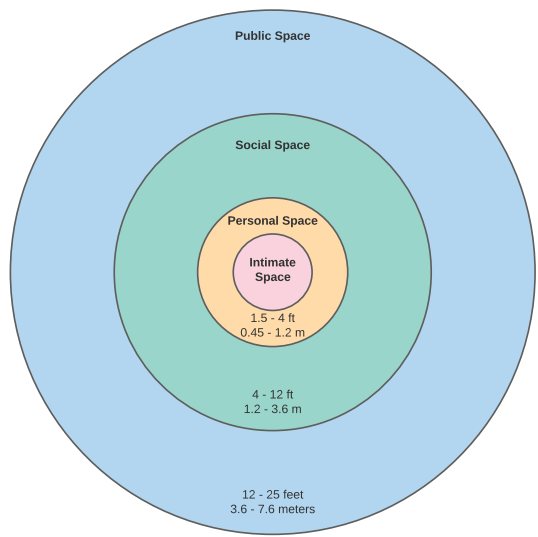

Fig. 7. Hall's taxonomy of social use of space, also known as proxemics [29].

## 6 WORKSHOPS AND FOCUS GROUPS WITH CHILDREN

Here we report two touch points with the target age group, that helped us to refine the design before our field validation.

First, before we had design mockups and in parallel to the design process, we used plushies chosen from among our positive mood board candidate images to engage children in informal interviews that explored the general design concept. We hosted two workshops with 6 children ages 7-11, that had five activities: 1) Children were asked what the creature should look like and not look like (using samples from our mood boards printed on cards); 2) Children were invited to elaborate on the idea that the creature appeared on their doorstep one dark and stormy night, using a storyboard format; 3) children chose one of the plush toys to enact how they might interact with the creature to calm and reassure it, and were asked to explain why they preferred that particular toy above the others; 4) children were asked to draw and explain where the creature might live in their home, and 5) children were asked how the story might end–would someone come to bring the creature home? Would it live with them forever? In terms of appearance, children were positive about the candidate mood board images and the range of sample plushies that we provided. They were intrigued by the story and glad to demonstrate how they might engage the creature once it arrived. Their playacting included a wide range of activities which blended elements of the creature signalling how it felt, and their own actions to calm it. In terms of the latter, hugging was popular, and some tickled the plushies as well. In terms of the former, children bounced the creature to indicate excitement, and talked about how it might signal its feelings with its face (including its ears). The children assigned the creature very different feelings and reactions in their playacting, which underscored the importance of leaving room for interpretation and free play in the engagement. This range of interpretation also helped clarify for us the importance of a very clear feedback loop to the child about how the creature was feeling in any given moment.

Next, after settling upon an initial design and creating the first functional prototype (figure 6), we sought children's initial responses to what we had created. Would they 'read' the creature's appearance as we expected? Would they respond to the touch affordances in the ways we imagined? Would the interaction 'game loop' be legible, motivating, and enjoyable for them? We conducted 3 focus groups with children ranging ages 4-12 (6-10 was the main age range, the extremes (4 or 12) were siblings) to address these questions. All were recruited from an area that was within the lowest 5 percent nationwide on the index of deprivation in the UK. Participants were a combination of White-British and BAME-British background (BAME stands for Black, Asian, and minority ethnic, a UK demographic). In each session, there was one sample creature passed from child to child (2 groups had 4 children, the other had 5. There were a total of 6 female and 7 male children among the groups). At first, children were allowed to explore the creature at their leisure, then parents or the experimenter would prompt them to try to "calm the creature". Initial exploration with the creature was fairly free-form, with only occasional prompting from adults, and guided hand-offs between children after 5 minutes of interaction. At the end of the session the experimenter led a 15-minute question and answer section with the group of children.

To explore the legibility of the creature's appearance, we looked at impromptu comments children made while engaging with the prototype, as well as comments during the interview portion of the sessions. Comments from children related to the creature's appearance included: 'He looks quite cute' and 'It looks like a kitten or something.' In terms of inviting tactile affordances, we looked at children's comments, as well as video coding their interactions with the prototype. Comments included 'It feels good' and 'It looks comfortable.'

We coded video of the sessions, tallying each interaction that happened within a 5-second period (we did not yet have the sensor data log function complete at this stage). If the same action lasted longer than 5 seconds, it would be counted as an additional instance of that action. The coding scheme was built from the bottom-up by examining the videos for common patterns [54]. The creature was considered to be cuddled if the child rested it against their chest, or cradled it in their arms. The creature was 'stroked' if scratched, petted, or rubbed actively in an area, rather than just held. This means that a creature could be held to the chest and scratched on the back at the same time, and both would be counted. The creature was considered to be shaken or thrown when rocked violently or thrown into the air. We originally also coded the number of interactions with the feet, but this action was conflated with 'waking' the creature up and couldn't be effectively separated from genuinely fidgeting with the feet.

| Child:              | 1A | 1B | 1C | 1D | 2A | 2B | 2C | 2D | 3A | 3B | 3C | 3D | 3E |
| Gender:             | M  | M  | M  | M  | F  | M  | F  | M  | F  | F  | M  | F  | F  |
|---------------------|----|----|----|----|----|----|----|----|----|----|----|----|----|
| **Cuddles creature** | 2  | 5  | 24 | 6  | 14 | 4  | 35 | 1  | 8  | 3  | 8  | 4  | 6  |
| **Strokes creature:** |    |    |    |    |    |    |    |    |    |    |    |    |    |
| -On head,back,sides | 3  | 3  | 9  | 12 | 20 | 5  | 49 | 11 | -  | 8  | 6  | 1  | 3  |
| -On belly,face      | 3  | 1  | 4  | 2  | 3  | -  | 11 | 1  | 2  | 3  | -  | -  | 1  |
| **Throws/shakes**   | 14 | 2  | 20 | -  | 31 | 2  | 8  | 1  | 1  | 8  | 1  | 2  | 6  |
| **Plays with:**     |    |    |    |    |    |    |    |    |    |    |    |    |    |
| -tail               | 1  | -  | 4  | -  | 2  | -  | 4  | -  | -  | 1  | -  | -  | -  |
| -ears               | 2  | 1  | 4  | -  | 1  | -  | 2  | 1  | 1  | -  | -  | -  | -  |

Table 1. Tally of natural interactions with early prototype during workshops. 4 children in 2 sessions, and 5 children in workshop 3.

The most common touch behavior by far was cuddling (110) and stroking (164). The children did manipulate the ears and tail, but far less than other touches, and not in a manner that suggested they were play-acting the creature's feelings. Another fairly frequent behavior was throwing or shaking the creature to 'wake it up'. This led us to realize that we needed to include some kind of motion sensor in the creature, that we could use to build in a negative haptic response to rough handling in the game loop, in future.

To explore whether the game loop was legible, motivating, and enjoyable–successful at eliciting ongoing interactions– we looked at spontaneous comments during play as well as the end-of-session interviews. In general, children were able to sense and respond to changes in the haptic feedback from the creature, and mapped these responses to the notion of the creature 'calming down' and 'going to sleep'. Their comments indicated understanding of the creature's states, for example: 'You're going to make him stressed... He don't like being stressed, he likes being happy' and 'He's stressed, he needs a nap.'

## 7 FROM RESEARCH PROTOTYPE TO FIRST RESEARCH PRODUCT

At this point, we planned to move to field study of the creature as a possible intervention. To get ready for this, we made an iteration of the creature that would: 1) tune its external design, touch affordances, and interaction mechanics and dynamics, based on what we observed in the focus groups, to maximize its effectiveness as an intervention to scaffold emotion regulation and 2) make it more robust for a planned in-home deployment–to move from a prototype to what Odom et al. characterize as a 'research product' [48].

The creature's appearance and role was quite legible to the children, so we did not make major changes as a result of their response to these aspects of the design. However, we did make some changes to the exterior design to address issues of durability for unsupervised, multi-day in-home use. Initially, the creature had a Velcro-closed bottom that could be opened to pull out and replace the battery (figure 6), but the placement of the access hatch still made it difficult to access the interior electronics for long-term care, while doing little to disguise the hatch from the children. So instead, we created an opening at the side of the creature that allowed for easier access to the core of the plush (figure 8). We sealed it with an invisible zipper that matched the creature's fur color to mask the location from immediate inspection.

Additionally the initial design of the creature had very little protection for the primary microcontroller stack, so a box was inserted around the core which provided some protection from both the poly-fill stuffing and diverted some of the impact force that accompanies squeezing (similar to how a cell phone case diverts impact pressure from your phone). The researcher could use this same opening to connect a USB cable to the microcontroller for easier battery charging and retrieving in-situ captured time-stamped data logs of interaction for post-deployment analysis.

We didn't have enough of the previous faux fur to make the 10 multiples needed, so we sourced a similar fur to make the multiples of the creature needed for an in-home deployment. Otherwise, body shape and other features remained essentially the same.

In terms of touch affordances, we made a number of changes. As we mentioned in section 6, children in the focus groups were not repositioning the tail and ears as we had imagined they might, and did not touch these areas of the creature much. We decided to try out an alternate design strategy for these extremities. We moved the circular Force Sensing Resistor (FSR) sensors that were initially in the creature's ears into its feet, positioning them between poly-bead fill to create a smooth rolling texture. The idea here was to offer this as a positive 'foot massage' style engagement with the creature. We moved the mechanical click buttons from the feet to the creature's ears, and removed the copper mesh in the ears and replaced this with soft sheet foam. The idea was that this interaction would feel more natural–the clicking embedded in the foam might feel more like manipulating an animal's cartilaginous ears. In the game loop, we recharacterized this as a negative behavior that agitated the creature. In the tail, we substituted the stiff wire placed originally to balance the creature for firm stuffing and a flex sensor that would enable us to sense when the tail was twisted or folded, which we also categorized as a negative behavior that the creature did not like, in the revised game loop.

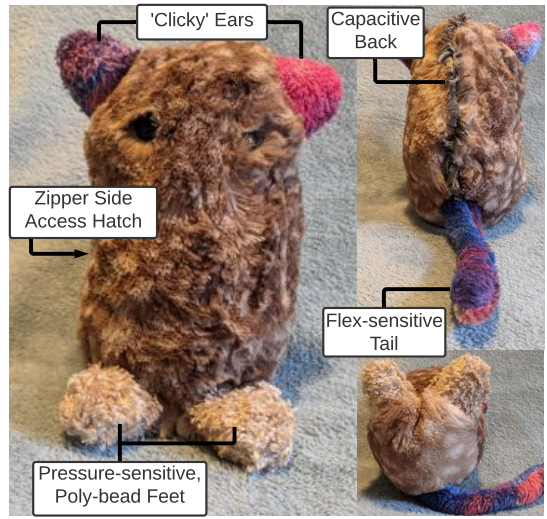

Fig. 8. Key features of the second iteration of Anxious Creature.

We found that children would hold the creature to their chest but were not often hugging the creature strongly enough for the pressure ball in the interior to detect anything, so we removed the pressure ball sensor in favor of additional stuffing. We noticed that the hugging was typically accompanied by stroking the creature's back, so we added a capacitive touch peripheral board to our controller stack. To integrate it well with the creature's short fur, we fringed the outward facing edges of a solid strip of the conductive fabric and inserted it into the center back seam, with a solid bare wire sandwiched along the backbone of the creature, stitched in on either side by the rest of the body fabric. The solid wire was then connected to the capacitive board. The visibility of the conductive fabric served also as another stimulus for the child to engage with. This made the material look and feel naturally part of the creature, keeping its exterior feeling soft and plush. It made detecting any petting/stroking motion on its back possible.

Finally, because we noticed children sometimes engaged in some aggressive shaking/tossing of the creature when waking it up, we added a gyroscope and used a moving average filter so that we could address these kinds of touches in the core game loop–abrupt motion would be perceived as negative by the creature.

In terms of the core game loop and interaction with the creature, our motor was originally controlled directly from microcontroller. For the research product version, we added a haptic controller (TI DRV2605) to make vibration pattern and timing easier to control and manipulate. Then, we recreated the basic heartbeat-to-purr haptic game loop for the in-home study, but with a few changes. First, we introduced the notion of negative touches–a rough shake, or playing with the ears or tail, could lead the creature to become more anxious. Whereas stroking, hugging, and foot massages calmed the creature down. Instead of a foot press to initiate the interaction, we embedded an on-off switch inside the creature's body, and put in a battery with much longer life, so that the creature could be sent home and remain on, only periodically needing to be recharged. This meant that we could use motion detection from the gyroscope to initiate interaction with the child. If the child woke the creature gently, it might begin in the purring mode, but if it was shaken, it would wake up anxious. This alteration meant that a child could nurture the creature and enjoy its company without it beginning in anxious mode every time. We made a number of small iterations to this revised game loop through internal testing before the in-home deployment, until the interaction seemed legible to our design team.

To summarize, the role and appearance were already legible to the children and so not changed much, but touch affordances and also the core game loop were amended based on the focus group results. We switched from a free play/play acting notion of how the ears and tail might be used, to a model of giving the creature some touch areas that led to negative responses, helping to shape the children's fidget patterns. And, we tuned the game loop accordingly.

After all of the changes were settled upon, we expanded production to 10 creatures, 8 of which were used for user testing, with 2 for in-house demonstration and maintenance rotation. Figure 9 shows multiples of the creature–note that the individual creatures had varying ear coloring, to help encourage children to see them as unique individuals, if they were exposed to more than one of the creatures initially.

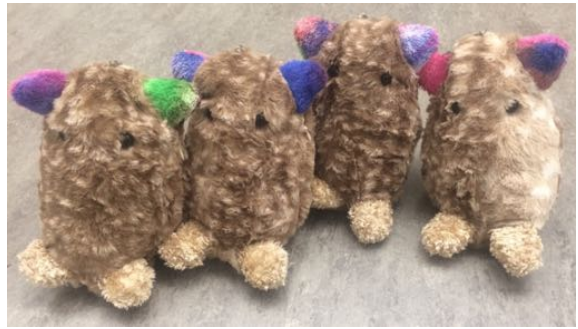

Fig. 9. Multiples of research product.

With this revised creature, we then conducted a smaller focus group-style interaction with two pairs of children, to ensure that the new details were legible and producing the responses that we intended. These sessions took about 20 minutes and happened in a quiet room without parents present. All children were approximately 8-10 years old. We used a simple evocative handout that took the form of a naturalist's partially completed notes (figure 10), with spaces for children to fill in their own notes as they investigated and interacted with the creature. The researcher would present the creature and these partial field notes to pairs of children, and allow them to interact with it naturally, before prompting them to help the creature calm down. Occasionally the researcher would intervene to shift possession of the toy from one child to the other. The researcher would jostle the creature occasionally so that both children would be able to experience a full anxious-to-calm cycle. Videos of the sessions was coded similarly to initial focus groups, with the addition of tracking foot interaction, as this was no longer conflated with waking the creature up. In these sessions, the children were observed to immediately attempt soothing behaviors on the creature. One child held the creature to their shoulder for the majority of the session without prompting, while another attempted to gently scratch the front and back of the creature for the majority of the session. Most children seemed to be hesitant to agitate the creature, avoiding contact with the ears and not shaking the creature. As with the first set of focus groups, all of these actions suggest that the design was working as desired–eliciting and rewarding caring, soothing behavior. When prompted by the experimenter to explain their understanding of the creature's vibration pattern, children talked about the 'fast heart rate' or 'cat-like purring,' suggesting that our haptic vocabulary was at least partially understood.

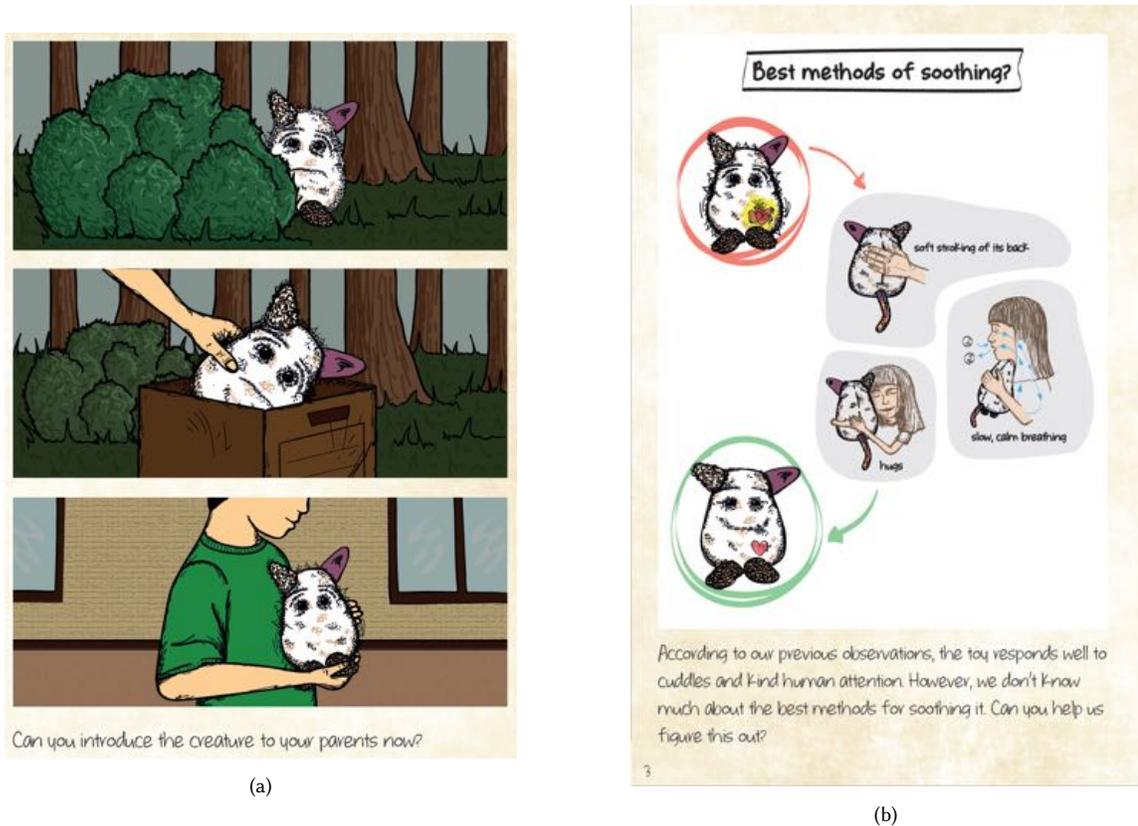

(a)

(b)

Fig. 10. Discovery book that accompanied the creature to home deployments.

## 8 VALIDATION OF FIRST RESEARCH PRODUCT

We then conducted 2 in-home deployment studies (total of 25 families, n=14 and n=11) to test the potential of the intervention with children for scaffolding emotion regulation. The results of these studies have already been reported elsewhere [60, 61], with particular emphasis on the psychological effects that the devices elicited in children and adults. However, the previously reports did not directly focus on how the observed psychological effects are related to participants' responses to the design characteristics of the creature (main focus of this section); nor how such empirical observations then contributed to the further design process (cf next section).

### 8.1 Procedure

We sent the creature home to families of a total of 25 children aged 6-8 (9 girls, 16 boys), across two subsequent studies: see [60, 61] for details. Families were recruited through schools in an under-privileged community in the UK. Each family kept the creature for 2-4 days for the first study with 14 families (depending upon school logistics and what day of the week the deployment started), and 7-8 days for the second study with the remaining 11 families. Alongside the device itself, the families were also given a discovery book (see figure 10) and a simple digital camera, which children used to take photos of and with the creature in their home, as guided by the discovery book. Within a day or two of handing the device back in, children were interviewed about the experience, with their school guidance counselor (study 1) or their parent (study 2) present. They had a chance to use the device during the discussion to aid their recollections, and photos they took during the home deployment were also used to aid their memory and help frame discussion. During the home deployments, the devices were keeping track of all touch traces on internal memory.

The Discovery Book was designed to provide a standalone version of the minimal backstory that we had crafted for the creature and told to focus group participants in person: that it was a lost creature that seemed to be in need of comforting.

## 8.2 Research Questions and Results

As was mentioned, the main results from these studies concerning the efficacy of the intervention have been reported elsewhere [60, 61]–in essence, children reported forming an emotional connection to the toy, and using it for emotion regulation. Here we briefly return to that data to highlight how the children and parents responded to specific design choices about the device itself, drawing on the same thematic analysis methodologies and process as described in the previous work. Questions we were interested in within this context thus included:

*8.2.1 Was the role legible and appealing to the children?* Across both deployments, nearly every child (23/25) named their toy and treated it as a living being that needed to be cared for, with feelings and mental states they seemed to take into consideration. The toy was readily adopted as a social partner, with children reporting they played games together, watched movies, engaged in pretend play, or slept in the same bed. In all these instances, the children were framing the experience as that of a partnership: the toy was actively involved in the activity; or transforming the experience by being close. As illustrated by the quotes below, most of the references to the creatures' 'emotions' have been directly linked to the interactivity (e.g., the 'heartbeat') and the projected impacts of child actions on the creature's emotional state or mood. In doing so, the majority of the children also built nests or other physical objects (e.g, clothes) for their creatures, to 'make sure' the creatures felt 'comfortable' and 'safe' in their new environment. Here are a few sample quotes from the interviews:

*First-Deployment Child 2*

I: And during the night? Did you sleep with him or was he in his nest? C: Oh. Actually, I put him in his nest when I got home and whenever I wanted to like rock him or stroke his belly, I would just take him out of the nest. But if he got all anxious, I would put him back in there and rub his belly whilst he's in it. I: Oh, okay. So, while he was in the nest you went to check on him and rock him. C: Yeah. I also discovered this new thing. Whenever you leave him alone and you're very far away from him, he actually gets really anxious. I: Oh really? So then you have to go back and help him calm down? C: Yeah. But I would actually never ever, ever, ever, ever leave him in his nest alone. I: Aw… So you carried him around with you then? C: Yeah.

*Second-Deployment Child 6*

C: My favourite thing to do with the creature was sitting on a tree with him, because… It made me really happy, because I can sit in my tree house -it's not really a tree house, I just sit on my tree that I was… I was sitting on there and talking with him. R: You were talking with him, what were you chatting about? C: Where did you come from and stuff. But then… I thought it came from a tree, that's why I put him up on the tree. R: Ah, that's why. So did he like it? C: Yeah. I didn't even touch him and he started purring. I think he likes the… feel of the tree.

*Second-Deployment Child 10*

R: And where did Missy live this week? Like where did she stay, where did she sleep..? C: Well, for the first two nights she slept in a pillow down there (pointing to a spot on the living room floor, next to an armchair). I had a blanket for her, but then I thought.. just to keep her a bit more warm, she could stay at my bed. R: Aw, that's so nice of you. Do you think she liked that? C: I think she did. Cos I've worked out that she needs to be warm to vibrate. R: Oh, does she? C: Well, usually when I'm doing this (demonstrating caressing the creature), keeping her close to me, I think it vibrates. And makes her happy, calms her down. R: So she likes being close to you? C: Yeah.

The interviews with the parents during the second deployment (n=11) further showed how the parents not only noticed the development of such caring relationship between the child and the toy, but also enjoyed and nurtured it. Several of the parents also described experiencing similar soothing effects themselves, again relating these to the interactivity and reactiveness of the prototypes.

*Second-Deployment Parent 7*

M: I think it was comforting for her… As I said, it was a lot like having her blankie when she's.. It's like a comfort… Yeah… When she came home from school, she'd grab hold of it and… Yeah. R: How do you think it worked to comfort her? What does it do? M: Um… It sort of soothes her when she's feeling stressed or… It just makes her feel safer I think. Yeah… R: And is it something specific about the creature that you think makes her feel safer? M: With <kid's name> I think it's the feeling of this, because it's a lot like what she's used to. I think the heartbeat and actual sleeping does make a difference because… It sort of… She'd be like "It's going fast, I better give it a cuddle". So yeah, she'd try to slow it down, try to make it relax again. I think that really made a difference. R: Mmm, so similar to her blankie but she also had to take care of it? M: Yeah…

*Second-Deployment Parent 4*

M: This is my super caring child! Yeah. He has treated him really really well. I'm quite surprised. R: (chuckles) You are? M: Yeah.. I mean, not in the sense that, <kid's name> is a really caring child, but in the sense that he's like now… making it pillows so it can sit (chuckles). You'd actually think he was alive.

*Second-Deployment Parent 5*

M: Yeah, I held it, and I tried to calm it down myself because I realised it got like.. how can I describe it now… a fast heartbeat… Um.. Like it's…like, trying to get the word. Like it's set to be that way, an angry creature or something. But as soon as I started stroking it, I realised it calmed down, not immediately, but after few strokes. And then, I realised it's also comforting for myself. And that helped me this morning. R: Oh! M: This morning.. when, umm.. because I had to send two off to school, and then the other two I had to get them ready for the interview, a meeting with a teacher. So this morning my stress level was really high. (chuckles) And I thought "oh my goodness!". And then the back door wasn't working, because the key didn't work. So… R: So you couldn't open the door? M: We couldn't open the gate. Anyway it was up and down this morning, and then I though "Oh! I don't want to be late for the kid's school and everything!". So I held the toy, and put it right to my chest and start to stroking it and I realised to my surprise that it helped him calm down as well. That was the very first time I felt it effectively helping me. Before that, I just played around with it, and not really realised how much it would have helped me… R: it wasn't in a hot moment, let's say, when you really needed to calm down, whereas this morning.. M: Yeah, this morning was.. I would put it as the point.. you know the main point, the biggest highlight. So it helped me as well.. do you have any for adults? (laughs)

*Second-Deployment Parent 2*

M: When we spoke about his emotions, and then when she was filling out the book, she said- Because one day she forgot him at home – so she said that because he wasn't with her, and alone, he'd be very sad! So, that day Coco was just sad the whole day. But she felt that she gave Coco comfort all the time, so he needed her basically. She didn't need him, he needed her (chuckles). [...] M: For example, if she felt, mmm, if she felt Coco, she would be alarmed. Like stroke him, he's not well. So straight away "Oh... Coco's beating, hold on!". So she would like, straight away: "Oh mummy, mummy! He's beating" or something. [...] M: I think it gave her like a sense of responsibility. "I need to look after Coco, and if I'm not there, Coco is going to be upset, so, I have to as soon as he... I have to calm him down". Yeah.

**Did the device evoke appropriate caring touch behaviors?** The high level of care and frequent touching behaviour has been a strong common thread across the interviews with children and parents as well as across the two studies. None of the parents or children reported any negative or violent behaviour toward the creatures; in fact, many children have specifically instructed others as to what kind of touch their creature likes, and made sure no one would 'hurt it'. In summary, 23 out of the 25 children have explicitly talked about a range of caring behaviours they exhibited with their creatures: 'cuddling', 'hugging', and 'stroking' were the most commonly used verbs, followed by kissing, keeping it close, rubbing, patting, or massaging (the feet). Many of the children 'discovered' a particular touch pattern that their creature 'liked the best'; interestingly, this sometimes involved interacting with the touch sensors in ways that we had not expected (e.g., placing the creature's back on the lap and stroking its tummy; or 'keeping her close and hugging'). This shows that although the children were oblivious to the placement of the sensors, the selected location and sensitivity enabled a range of touch behaviours that were still legible and evoking meaningful interactivity without turning the interaction into a overly simplistic gameplay (e.g., the creature reacts to stroking of the back, but nothing else). The quotes below were selected to illustrate the range of caring touch behaviours encouraged by the design, as well as show that these could also take place in addition to another activity that was done 'together' with the toy: reading, watching something, playing a game, or falling asleep.

*First-Deployment Child 2*

I: Aw. And before you said you watched TV together... Before then, did you play with him at all? C: Eh... Well, I did rock him a little bit. And I kept on rubbing his belly to see if he was calm or anxious. I: Oh. And how could you tell? What happened? C: Well, once you like shook him he went all anxious and he was like showing to the interviewer. I: Aw, his heart was pounding really fast. C: Yeah. And once I started rubbing his belly it started to slow down, and then he started purring. I: Okay! So, you rocked him a bit, you rubbed his belly, you watched TV with him and then you went to bed... C: And still watched TV. I: Still watched TV. And then he was sleeping and purring. C: Yeah. And then I went to sleep. And then my mum went to sleep. And then my sister went to sleep.

*Second-Deployment Parent 2*

M: So, stroking, hugs, always close! So she had him between her legs when she's doing something. And then when she feels it, she goes like... hug him, and put it close to her chest. [....] M: For example, if she felt, mmm, if she felt Coco, she would be alarmed. Like stroke him, he's not well. So straight away "Oh... Coco's beating, hold on!". So she would like, straight away: "Oh mummy, mummy! He's beating" or something.

*Second-Deployment Child 9*

C: I usually holded (sic) her like close to me and stroked her down her back. I: Mhm. So did that work to calm her down? C: Yeah. She started purring in no time! [...] I: What was your favourite thing about having Wootie with you this week? C: My favourite thing about having Wootie with me this week was because I got to cuddle and stroke something.. [...] C: I felt best when the creature and I'... were cuddling and stroking each other because... I find it really relaxing and I get relaxed quicker when I cuddle and stroke.

*8.2.2 Was it engaged with frequently?* As can be seen in figure 11 the trace logs from the toys during both deployments show regular interaction with the toy. In particular, children seemed to interact most regularly with the back of the creature, suggesting that hugging or stroking occurred on a regular basis. It was also notable in the data that even during long periods overnight and during periods when no other sensors were active, the tail sensor was often triggered suggesting that the sensor itself was reading false positives. We have removed the tail data from the diagrams shown in figure 11 as a result, and the design team noted this for the future revisions as something that needed addressing.

During the field trials children often noted family members playing with the creature, so it's possible that this data reflects multiple people in the household, but even that shows us that the creature is facilitating interaction. For more details about individual children see our analysis papers [60, 61].

*8.2.3 Were the creature's responses in the 'game loop' legible and appealing?* As already indicated in the first section of the findings, both children and parents projected emotions onto the devices, saw their interactions with the device as meaningful and impactful (in terms of affecting the state of the device), and did not report any inconsistencies they would notice. We argue that it was likely the ongoing legibility and stability of individual interactions which enabled the development of the broader caring relationship: the game loop was understandable and seen as consistent over time (illustrated by the myriad statements in the format of "my creature likes <this>, but not <this>"); it was not overly simplistic (as illustrated by the intricate stories and emotional projection both children and parents reported above and in prior work [60, 61]), and finally, it was appealing (as illustrated by both the ongoing interaction as well as observations from children and parents describing the feelings of happiness or calm during the interactions with the toy).

*Second-Deployment Child 12*

R: But the creature was happy? C: Yeah. R: Why was he happy? C: Cos I was mainly hugging it a lot! It kept me happy as well.

*Second-Deployment Child 10*

R: Did you discover anything else? You said you think she responds well to holding her close.. C: I like this.. I also like keeping it under a blanket and lying down with it next to me, covering it.. I also like this (demonstrating). R: So you're holding it close to your tummy and then you're leaning over it. C: So.. Close, yes. Usually lean over it.. Sometimes I stroke it, sometimes I keep it in the covers and keep it close to my tummy.

*Second-Deployment Child 6*

R: And what was your favourite thing about having the Creature this week? C: That we can do loads of stuff. We can do this, and do this if he just keeps purring and you want him to get mad and then make him purr again. R: So you can cuddle with it, you can press his ears to make him mad, and then you can calm him down, so you did that? C: Yeah. R: So you did that? Why was that? Did you want to calm him down sometimes? C: I like calming him down.. because when he's just purring it's just... it makes me calm.

A further—if unintended—support for these observations about the relationship between the moment-to-moment game loop and resulting long term engagement came from a failed deployment during the second study. Here, we mistakenly sent out a prototype with the very early gameplay version (prior to any of the gameplay tweaks). The core gameplay loop was very similar to the other units in terms of the length and types of interactions that 'soothed' the creature as well as the vibration patterns of individual creature states. However, the sensitivity of the gyroscope and the tail sensors was much higher than in the other units: the gyro input was not reduced when the creature was 'happy' and the tail sensors were still active, which led to frequent 'frights' for the creature during even the slightest

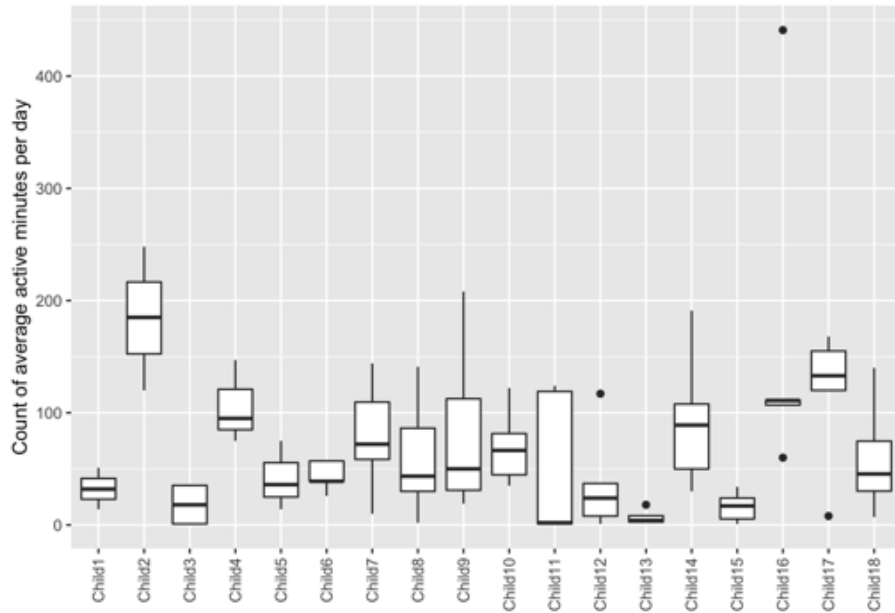

(a) Daily active minutes per day during the first field deployment.

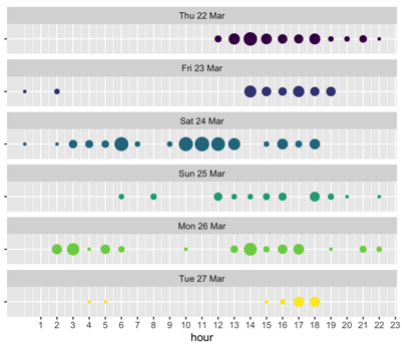

(b) Child 14 hourly activity count over deployment

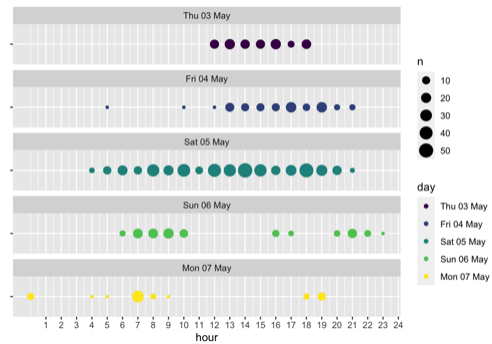

(c) Child 16 hourly activity count over deployment

Fig. 11. Touch trace counts per child from the first deployment. With two data breakouts provided for Child 14 (11b) and 16 (11c).

movements. The child was the only one who did not keep on using the toy; her interaction patterns started similar to all the others, but then dropped off after approximately 3 days. As illustrated by the interview snippet with the child's parents, the resulting 'neediness' of the creature and the inability to close the loop with a stable positive state (purring/sleep) has soon led to frustration and broke the relationship and emotional projections that has been evident in the other interviews.

<Parent interview> R: And what did you think of it? M: It was okay.. I think that as soon as you touch it, the heartbeat starts so it just feels like it's constantly beating. There's no sort of- you can't hold it with a break, if that makes any sense. It will purr and then literally like in 2 seconds it will start beating again. R: Really? That's strange. Because it's supposed to calm down, purr, go to sleep... And then after it takes quite a while before it wakes up again. M: As soon as you touch it, it will start! R: Okay! That's interesting.. What happened while <kid's name> had the creature with her? What did you see her do? M: Mainly cuddle it. And she did stroke it quite a lot. F: She liked playing with the things on the back, the... She found that quite therapeutic. She was... doing that a lot. M: But the last few days she just hasn't be bothered with it all, it has just been stuck by the telly where she took her last photo. (chuckles) That's it. [...] M: The first few days she was quite pleased, she wanted to show everybody. She wanted to take it to his mum's and show her... And she took it to her friends' house to show them. She was quite happy with it. But then all of the sudden, she just stopped. Even though I did try and like push her towards it, she wasn't...

### 8.3 Limitations

From a design intervention point of view, this first research product was a success–the device affordances were legible and had the effects we intended, and the intervention strategy showed great promise. However, as we mentioned in the introduction, the gold standard for wellbeing interventions in the health field is really controlled trials with a far greater number of participants than we had in this initial home deployment. Yet the team was stretched to achieve this device deployment to 25 families–we only had produced 8 units that could be used for research, so we had to rotate them among families. To create even that many devices took many hours of researcher time–one of the students felt she'd turned her home into a sweat shop where she was sewing for hours. And, the devices were fragile–the haptic motors started to fail and needed to be replaced. Using larger batteries helped with deployments, but could not last longer than 3-4 days and were cumbersome to replace (something which wasn't realistic to do by parents). As a result accomplishing even week-long deployments required the research assistant to physically travel to participants' homes to replace batteries. Transferring data from the devices was also cumbersome and time consuming. We simply were not ready to scale up to running an RCT with 120+ subjects over a period of 4+ weeks; as in the numbers necessary for what would be still considered a very modest RCT design in mental health [10, 17].

## 9 FROM RESEARCH PRODUCT TO COMMERCIAL PARTNER

Fortunately in our case, the non-profit that helped to fund the initial device design, development and research, was excited by the potential shown in this initial study, and was very interested in advancing the project forward toward an eventual commercial release. They saw this device as something that could complement their existing emotion regulation curriculum for schools, providing a child-led and situated intervention that could be used by students in the home; thus addressing one of the key issues across SEL programs – transfer of interventions from school to homes. They brought a new partner into the project–a product development company with an extensive background developing health-related socially assistive robots, that was very interested in evidence-based design.

With this new partner in the mix, we could work together to produce an even more robust research product that we could use for RCTs and a wide range of research contexts, along the way to the product company developing their eventual commercial product.

In essence, this newly formed larger team had two parallel objectives: 1) creating a robust research product for conducting continued research into the efficacy of the intervention and 2) creating a commercial product prototype that would successfully appeal to markets the company identified. The latter objective introduced what Ko et al. [6] term 'adoption-focused design' into the process. At the foundation of both objectives was the importance of developing a research-validated intervention (valuable to all parties). Beyond this, there were slightly different considerations for each: for the research product, we were primarily interested in introducing robustness for field trials. We also wanted to keep the possibility space open for further development, so we wanted flexibility in the underlying platform. The commercial product focus was on price point and market appeal (as per Ko et al. 2015 [6]).

### 9.1 Translation Process

The commercial partner had their own design team, who would be working to take lessons learned from the creation of the initial research device toward translating them into a commercial product prototype that would address identified markets. What follows is an overview of how the research design team shared design insights and worked with the commercial designers to refine the new device.

*9.1.1 First Steps–Sharing Research and Design Materials, Agreeing on Process.* Remote kickoff calls were begun with three key objectives: outlining goals and objectives for a minimum viable product (MVP) pilot, defining how to collaborate on ongoing development, and discussing key learnings from work to date. Priorities from the researcher side included: 1) Accurately conveying the core design values and learnings from the research to date. 2) Helping the product development company to retain the essence of the research prototype/product in terms of the efficacy of the intervention, as they adapted it for robust production and for markets they identified. 3) Ensuring that the resulting work would support further research at scale.

Priorities from the product development company side included: 1) Understanding the core mechanism and efficacy of the intervention, and learning about the process taken so far. 2) Identifying or stratifying development into areas that were clearly defined and opportunities for further exploration. 3) Understanding existing design language and values, for effective and cohesive collaboration. 4) Aligning and agreeing on the process for further development.

*9.1.2 Project Kickoff–Onsite Workshop with Product Development Company.* Next, the research team attended a two-day on-site workshop led by the company at their location. At this stage, the company was driving the process. They circulated notes ahead of the meeting which included a Theory of Change, and Design Intent (figure 12), distilled from the prior remote discussions. The first day of the workshop was devoted to clarification and discussion of goals for the project (producing an MVP: minimum viable product, that could also serve as a research product), and diving into details of design intent and gameplay. Then, the company did a show and tell of their initial work on ideas for where the design of the product would go. They shared sketches (figure 14) and also urethane foam mockups of potential shapes for the creature (figure 15).

The company's concepts built upon some core features of the research prototype design, and iterated upon others. For example, the new designs included a rounded body, and key body features such as ears and feet. Some sketches showed tails, and the research team shared what was learned about false signals from the tail, which led the team to decide upon a smaller tail that did not include a flex/bend sensor. Like the research prototype, the company's mockups focused on the intimate rather than interpersonal zone of interaction–the new mockups did not have mouths nor moveable eyes, and were intended to be picked up and held close by the child.

One interesting extended conversation that the combined team had, was about the base posture or pose of the creature. As can be seen from figure 15, the company tried out prone positions for the creature as well as the original upright pose from the research. In discussion (and confirmed with the company's tests of the urethane models with children), we realized that the more neutral and ambiguous pose allowed the child to more flexibly read the creature's state based on haptic feedback. The creature could seem guarded or could seem happy, based on anxious heartbeat or purring, with a neutral pose. If the creature was always in a limbs-extended, relaxed pose, this might undermine the legibility of the core gameloop. The research team also noticed that the early sketches seemed to drift into a more childlike set of features, and we encouraged the company to remember that the creature needed to also exude a sense of potential strength and self-efficacy. Here we note that it was incredibly valuable to get the team together in-person, for extended discussions that included the handling and comparison of physical models. Team members picked up mockups to see how it felt to hold them close, and could view them in different positions and think in an embodied way about how they would be experienced by children. We feel this was a crucial stage in the transfer of knowledge between the teams.

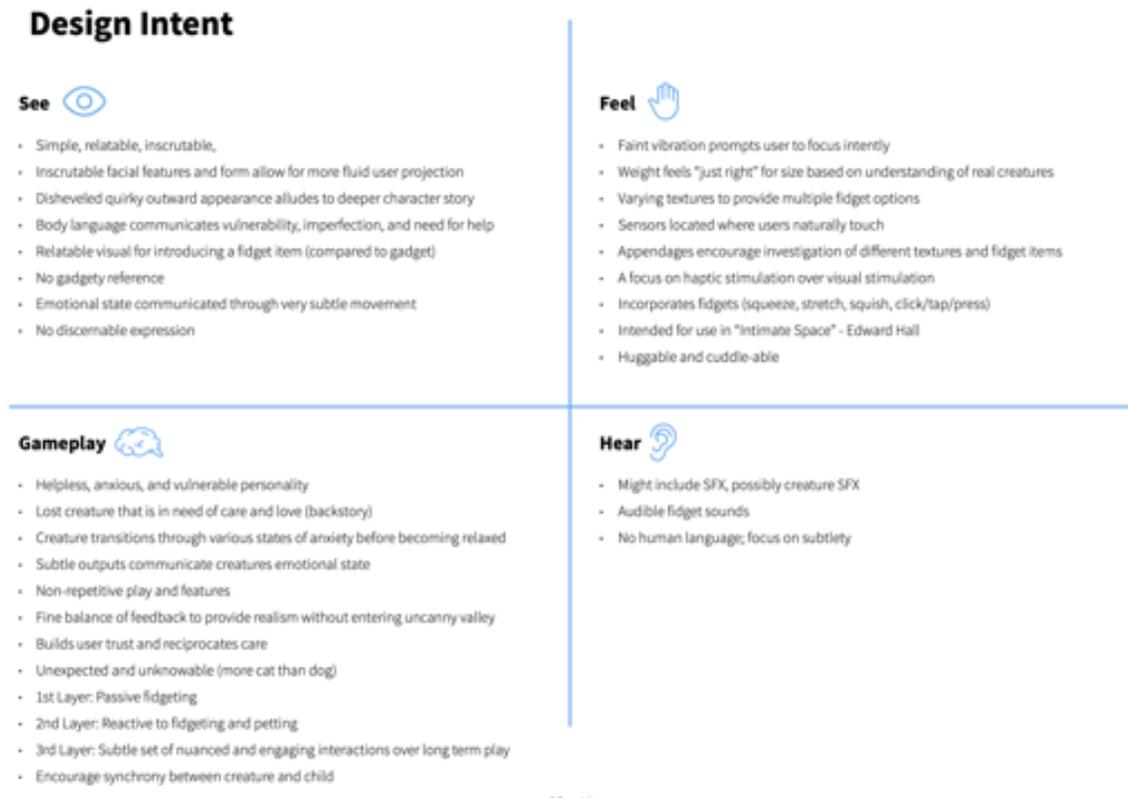

Fig. 12. Design Intent slide from project kickoff meeting.

During the workshop, we also spent quite a bit of time discussing two potential markets for the resulting product: schools and home users. Each had different priorities and goals, and existing constraints. As Ko et al. 2015 [6] have noted, considerations of market are a crucial part of moving from research prototypes into adoption of robust and viable products, and the research team gained useful insight into this process, that will serve us as we continue to study how these interventions may move to viability.

*9.1.3 The New, Translated Design.* The company spent some months working from this kickoff meeting, to create the product prototype. In figure 16, we show the company's product prototype. The company modulated the design to look more like a real-world, albeit ambiguous creature. The fur is a more neutral color. The company's creature looks a bit more childlike, with a rounder belly and wideset eyes. One interesting carryover is that the company worked very hard to keep a wild, individual quality to the creatures, by paying close attention to how they sourced the plush. They chose a plush that had subtle variations in color and pattern, staying away from more uniform and cheaper fabrics based on our advice from the workshop.

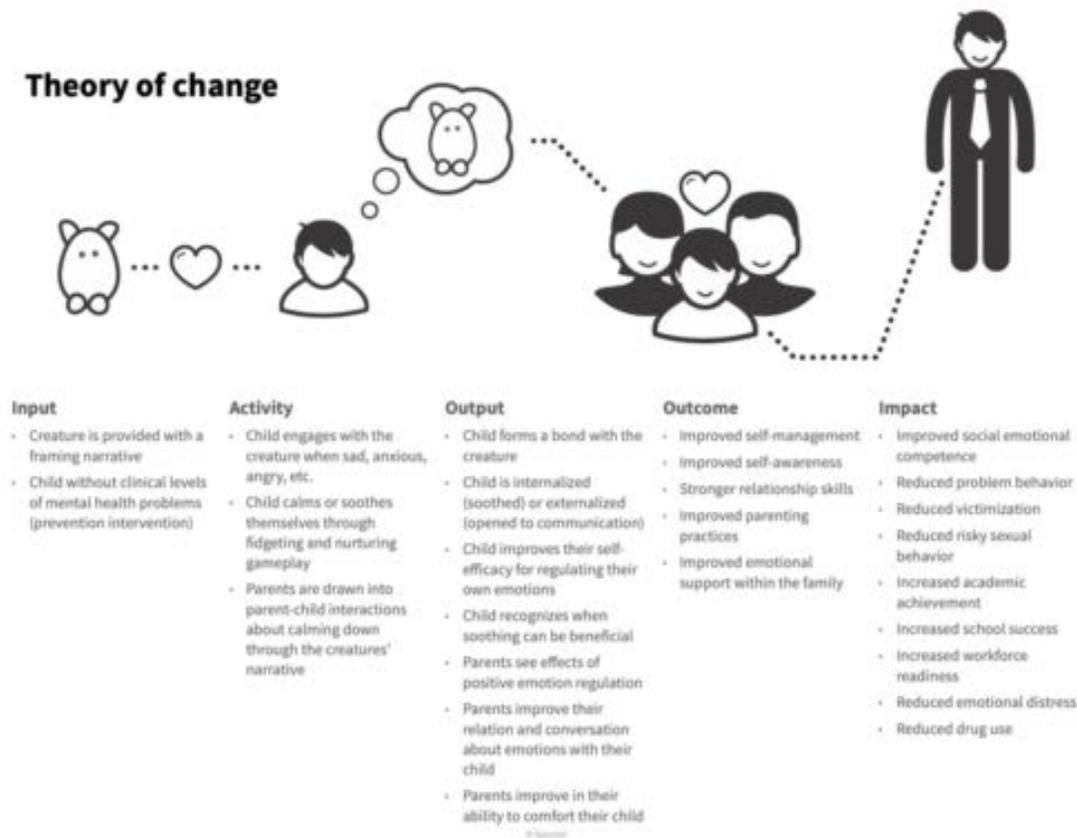

Fig. 13. Theory of Change slide from project kickoff meeting.

The sensor placements in the creature were similar but not identical–the child could play with the feet and the ears, as well as a small tail. Initially all three had sensors (manipulating clickable buttons in the feet and ears produced positive response in the creature, and pulling the tail negative response), but the final production version removed sensors from the feet and tail, with non-electronic beanbags added to the feet. Children could also hug the creature to influence how it felt. Instead of using conductive fabric on the back, the company used capacitive sensors internal to the device to detect hugs and stroking. As with the research prototype, the company included motion detection so that the creature knew if it was picked up, and handled roughly versus gently.

An important shift that the company made in terms of feedback to the child, was to replace the onboard motor haptics with a sound speaker-based response from the creature. The haptic motors were an ongoing failure point during the research deployment of the prototype creatures–we had to replace many broken ones. They were also relatively expensive. The company knew of sound speakers that created vibration that could mimic a motor's haptic feedback, and these were used instead.

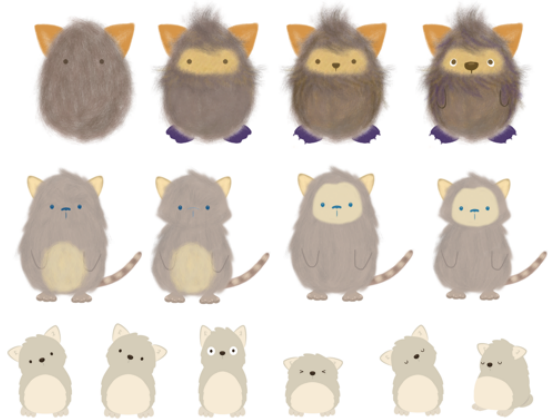

Fig. 14. Initial sketches of character from Partner.

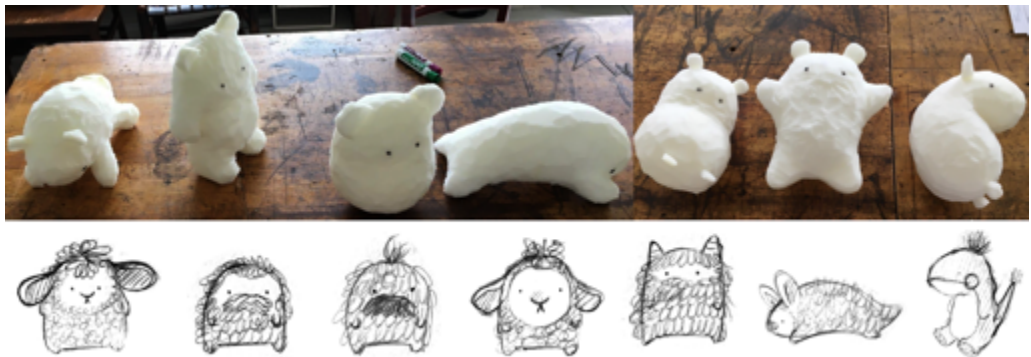

Fig. 15. Initial sketches and sculptures of character shape and pose from Partner.

The company also spent a great deal of time elaborating and refining the core gameplay loop for the creature. They kept the fundamental cycle: the creature would become 'anxious' and show this through a rapid heartbeat, and could move into a calm happy state, shown with a purr. However, the company added some more subtleties and modulations. For example, they added gentle sounds made by the creature both initially, and also, in response to touch, that vary according to the creature's heart beat.

*9.1.4 Tuning the Interaction.* Once the company had built the new product prototype, they distributed initial copies to the research team, and we collectively engaged in a series of iterations based on engaging with these prototypes and with subsequent updates to both hardware and software. Alongside this ongoing dialog, the company was also introducing the device to people in their target markets, collecting their feedback as well, and using it to make adjustments.

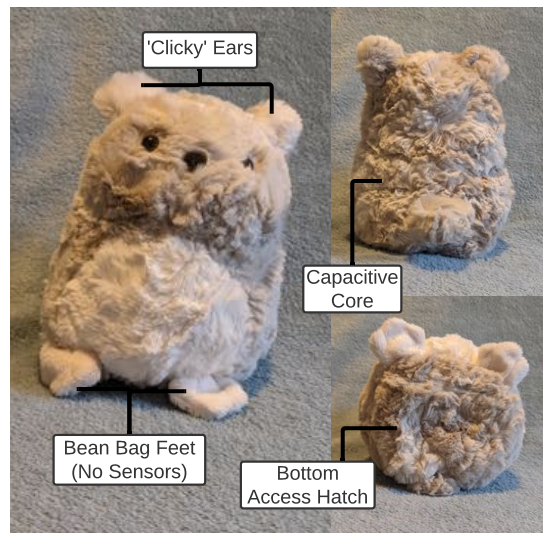

Fig. 16. Diagram showing key features of the final commercial design.

One area where adjustments were made was in refining the balance of the sounds and the haptic feedback. The company ended up including a switch that allowed end users to modulate how much vibration was part of the interaction–'low mode' had less, and 'high mode' had more.

The hardware had to be adjusted as well, because of issues introduced by swapping the conductive fabric that was part of the research prototype (figure 8) for capacitive touch sensors. The sensitivity and placement of the first set of capacitive touch sensors led to the device not picking up on the full range of hugs and pats from users, which then caused their mental model of the game loop to fail. Also, some who picked up and held the first prototype didn't like the hard, ungiving sensation of the internal frame. The company made changes to make the device feel softer, and to increase the responsiveness of the sensors to key types of touch from users. Key changes focused on tuning the touch tracking, to create a consistent game loop that accommodates different patterns of petting and holding the creature.

Finally, the company made many small, subtle tweaks to the core game loop that were software based, and thus easier to change. During this stage, the company would frequently circulate new versions of the code base for the researchers to test out. Both teams also had potential end users and their caregivers try out the interactions to check for their legibility, in moving toward the final release version.

### 9.2 Validating the Translated Device

The full research/company team did not engage in a formal comparison test of the research prototype versus the product prototype, but we did collect reports and insights across a variety of sources that help verify that the company's product design replicated the core interventional aims of the research prototype.

*9.2.1 Qualitative data stream: long term deployment with a school counsellor.* Here we share material from interviews with a counselor who worked with the team to deploy both device versions with children, toward demonstrating the success of the product prototype in creating similar effects. This is a school counselor working with primary school children (so aged 5-11), who collaborated with the research team over a period of more than a year. She has served as our key contact with the school, helping coordinate and recruit other teachers to use research and product prototypes in the classrooms (which had been subsequently paused due to the pandemic); as well as using both the initial research prototype (approx 6 months), and the commercial design (approx 4 months) within her clinical sessions with children at school.

The counselor noted that children were 'caring and nurturing' to the new model, and that they 'cradle and sooth' it–both behaviors that we witnessed with the research prototype. The counselor noted that 'it seems to sooth the children and calm them. It looks like it gives them comfort.' When children were upset, the counselor noted that 'there's definitely the sensory, the sort of stroking them, looking down at them, and then they might be too absorbed in their own.. whatever it is that's going on. That they certainly, the touch thing where they're holding it, hugging it, stroking thing, it seems to be the comfort thing.' Also: 'it seems to be very, very soothing, an immediate thing.'

What about the changes from the research prototype to the commercial design? The modification of the tail? The more baby-like features? And the addition of sounds? The counselor reported enhanced value from the sounds the new creature made: 'they were charmed that it made noises and was talking to them' and that children are 'definitely talking more with this one.' The counselor noted 'they speak to it and listen for a response' particularly when they were 'calmer, not in such an upset headspace, they might say things like 'Oh, it's talking to me! What are you saying to me?', 'Oh, it's saying it's hungry or it needs a cuddle', 'Oh, it's saying it wants to talk to me'.. So there's lots of that. And I do feel that it's possibly because of the sounds. So, you know, they'll sit there and when they hear a sound, it's like there's something going on between the creature and the child.'

Concerning the rounder shape, more baby-like features, and shortened tail of the commercial units, the counselor shared that children were 'much more delicate with these ones. I think partly that's [chuckles] to do with the tail. Because the tail was quite tempting to pick it up and swing it round. And, again, personally, I feel that it's a much more delicate-seeming creature, so they're much more careful with this one.' The counselor also noted on first impression that 'this one appeared more fragile.'

Overall, this counselor's impressions suggest that the core design choices and intentions (role clarity, touch affordances, and interaction legibility) carried through in the commercial design, and that the modifications the commercial team made added enhanced value to the experience for children.

*9.2.2 Mixed methods: Replicated at-home deployments with 20 families.* Following the positive responses from the school counselor, we have replicated the at-home deployments with 20 families of children aged 8-10 years, who have not been diagnosed with a mental health condition or a developmental disability.

The families were recruited through a post on online support network for mums (Mumsnet) around May 15th, and received the toy through post as UK was on a nation-wide lockdown at the time. Primary target of the deployments was to observe whether analogous themes arise in terms of usage and perceived efficacy on child's emotion regulation, and to validate the team's ability to collect outcome data remotely from both parents and children, as a preparation for an RCT trial (with no statistical analysis planned, as is common for feasibility studies in mental health due to underpowered sample size).

Using early product units, we have replicated the final packaging by hand and shipped the units to participants (see figure 17), with the data collection happening online through questionnaires at baseline (2 weeks before deployment), 2-4 days before deployment, and 2 weeks post deployment. The questionnaires were a combination of established outcome measures (such as the Strengths and Difficulties Questionnaire [26]) and open-ended questions mimicking key lines of inquiry from prior studies with research prototypes.

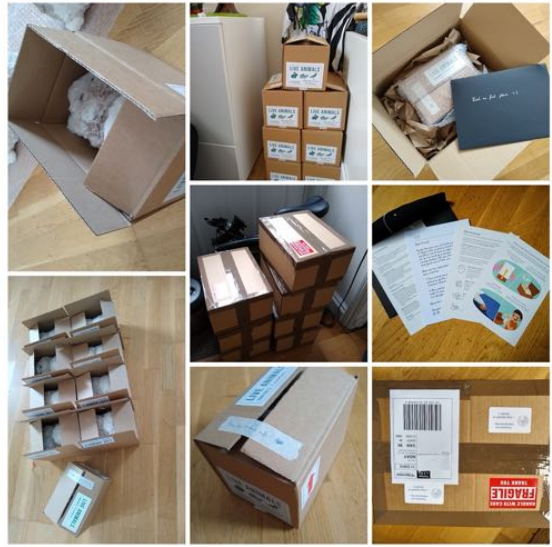

Fig. 17. Collage showing the packages and components sent out to the feasibility study families.

The results show that parents and children were similarly engaged and supportive of the commercial prototypes as with the prior research prototypes. Specifically, 19 out of the 20 parents (95%) thought that *the toy was helpful for their child*, and almost all reported that *the toy helped their child calm down when they needed to*. 16 out of the 20 parents (80%) reported that their child had *difficulties with emotions/concentration/behaviour/being able to get on with other people* at the start of the study (baseline assessment), as measured by the SDQ. After having the toy for two weeks, 13 out of those 16 parents (81.3%) reported that their *child's difficulties were better since receiving the toy* and thought that *the toy played a positive role in that change*; the remaining three parents said the *difficulties were still about the same*.

For example, parents have mentioned that:

- "In moments of high emotion, it is difficult for her to reason. Picking up the toy, calming down then allows her to be able to listen and then problem-solve" [Mum of 9-year-old girl]
- "He says he likes having an extra thing to calm down. He has enjoyed playing with it and soothing it. [...] He's enjoyed nurturing it. We've talked more about feelings. It was something to look forward to during a really boring time when we couldn't go out." [Mum of 9-year-old boy]
- "It has given him a separate focus point and a way of calming down without being told to, he will hunt the toy out when stressed and use it to calm down. He has chewed less (he chews when anxious) since having they toy, even though he doesn't use the toy everyday." [Mum of 8-year-old boy]
- "The toy (named Rebbie) has opened daily conversations about how the toy is feeling,looking after him, checking and calming him down. [My son] had one occasion where he was very upset. I gave him the toy to stroke and it calmed him down immediately and we were able to talk about what had upset him. I found what would been a stressful situation completely stress-free for me, it was wonderful to have something that calmed him down and help him feel better. [My son] has loved looking after the toy." [Mum of 10-year-old boy]

Interestingly, 18 out of the 20 parents (90%) thought the toy had been helpful in other ways as well, such as facilitating discussions around emotions and how to deal with them, siblings/parents using the toy too and finding it soothing, leading to a more peaceful home due to fewer temper tantrums, etc.

*9.2.3 Observational: Tracking uptake in the market.* Finally, the ability to observe how commercial units are taken up and reviewed outside of study settings was a crucial part of understanding the potential impact and appropriation in real-world settings. The commercial units were available for sale on Amazon and on the commercial team's own website from October 2020, and all available units (approximately 9500) were sold out by early March 2021. The Amazon ratings of the product remained consistently high throughout this period (in between 4.6-4.8 out of 5), and the text reviews showcased analogous parent and child engagement themes as seen in our prior studies both with research and commercial units (see figure 18 for examples and average ratings).

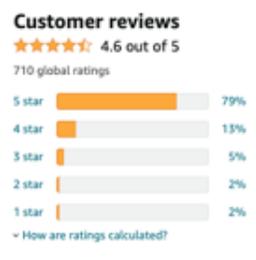

Fig. 18. Overall rating and 3 Sample reviews from the commercial product page on Amazon reflecting our expected affordances.

While much further work is required to understand and track the efficacy of the resulting commercial units, the data so far is indicative of both the translation being successful, but also the observed engagement and perceived effects likely transferring from constrained study settings also into the real-world (which is often challenging in the context of digital mental health interventions, cf., [44, 63]).

### 9.3 Reflections on the Translation Process

Overall, the research team was very gratified by how the commercial team adopted and then evolved the design of the creature. Here we provide a few reflections and recommendations for others who might want to work with a commercial team to create a more robust and scaleable version of their work toward creating multiples for larger scale testing. 1) **Aligned values**. It helped in this collaboration that the company already was very interested in the domain of socially assistive robots, and had prior experience building robots for other health contexts. This meant that they were not trying to turn the prototype into a typical toy, but rather deeply understood the core design ideas. If possible, it is a great idea to find a commercial partner that already has relevant knowledge and experience. 2) **Ongoing communication and artifact sharing**. It was important to the design translation process that we had ongoing communication, and that we shared artifacts both early in the process, and in the ongoing tuning phase. The in-person kickoff at the company allowed rich dialog about the nuances of the original design, and helped the company to internalize key design values and build upon the original design thoughtfully. The sharing of the physical prototypes, and then the code updates, helped everyone to grasp and give feedback about the design evolution. One really could not understand changes in the core game loop and haptic cues without feeling them for oneself. Overall, the artifacts were very important as communication points (as per Remy et al. 2015 [50] and Concalves et al. 2011 [25], as well as Gaver and Bowers [3, 19, 21]. 3) **Recognizing and adapting to divergent needs**. As we mentioned, the company was working on understanding possible markets and evolving the design to meet those needs, while the research team wanted a device that could support ongoing research. In the process, we realized that certain key features of an ideal long term research product needed to be jettisoned from the commercial prototype, in the interest of cost management and durability. In particular, the ability to collect time-stamped touch trace data needed to be removed, to avoid the necessity of a secondary battery and additional system complexity. Also, specific appearance, affordance, and game loop decisions got made on the commercial path that we might want to vary and experiment with over time in further research studies with different end user populations and contexts.

The research team engaged in extended conversations with the commercial team about how we might create a divergent research toolkit alongside their path to product. This would allow us to continue to conduct research, which would in the long term benefit the non-profit and the company as well, given their evidence-based design focus. The commercial development team agreed to provide instructions and components to the research team, so that we could develop a parallel, 3d-printed, Arduino-driven set of research prototypes to continue to do our work. The research team has used these instructions and components to successfully build devices that we can use as a testbed for further iteration of all of the relevant design variables in the intervention–role and relationship (by changing 'skins' of the device); touch affordances (by adding and changing sensors), and interactions (by modifying the game loop and behaviors freely by recoding the arduino core of the device). This will allow us to engage in a longer term research agenda that takes advantage of the increased durability and scaleability of the commercial design, toward future RCT-style trials as needed.

## 10 REFINING THE DESIGN SPACE OF INTIMATE-SPACE SOCIALLY ASSISTIVE ROBOTS

The translation process also helped us to more clearly articulate the notion of 'intimate space' socially assistive robots. Based on this translation, we would define such robots in the following way: well designed intimate-space SARs drive interaction in the intimate zone by: 1) evoking for the user a persona/role that is appropriate for evoking intimate-space interaction, 2) providing touch affordances (and an overall form factor) that facilitate intimate-space interaction, 3) providing a feedback system that structures and rewards intimate-space interaction.

To help clarify this design space, here we consider exemplars that we found in reviewing the literature around tangibles/haptics and socially assistive robotics. In the following table, we briefly consider examples of both research and commercial devices we would consider to be in this category to some degree: Paro [38, 52], Haptic Creature [5, 69, 70], Huggable [40], Qoobo [18], as well our Anxious Creature (figure 8) which became the final Commercial Ready Design (figure 16). For each, we briefly characterize the persona/role of the device, touch and other affordances, and the interaction loop.

As outlined in the Design Process, section 4, our team focused on three key principles during the construction and evaluation cycle, 1) Projecting a persona/relationship, 2) Creating inviting tactile affordances and 3) Eliciting ongoing interactions. Our overall goal was to create a device that would encourage a child to feel comfortable enough to integrate the creature into their intimate interaction circle, toward scaffolding self-soothing.

We settled on some design features that we see in these other robots/SARs, that we believe encourage the end user to welcome the robot into the intimate space of interaction. First of all, the robot is of a small size. All of the robots in the table are easily lifted and carried, and can readily be placed on the lap. At this size, it is convenient, and also nonthreatening, to bring the device very close for petting and hugging. Considering projecting a persona/social role, this size factor works well with the personas/roles that were chosen for each of these robots as well. All of the robots take on the form of an smallish animal that a person would feel comfortable caring for and connecting with. Interestingly, other than Paro, the design choices tend toward abstracted versions of creatures rather than evoking specific creatures.

Let's now consider inviting tactile affordances. All of the examples we include have soft surfaces, encouraging touch. In addition, all have some form of active feedback mechanisms that encourage touch, whether mechatronic movement, haptic vibration, sound, or some combination thereof. Huggable stands out as also providing affordances and feedback mechanisms that are more appropriate to the personal/conversational zone of interaction, with its capacity for conversing in words (vs. animal noises) and its directable gaze. We consider Huggable to be bridging between the intimate and personal space zones of interaction.

In terms of eliciting ongoing interactions, the robots in our table vary in their interaction loops, but a common theme is the use of simplified, stylized versions of responses of domesticated or harmless/baby animals to stimulus and connection. This is true for Paro, Qoobo, and Haptic Creature, as well as for our research prototype and final commercial design. The interaction loops in all of these robots emphasize eliciting and then responding to touch by the user, forging a positive connection and leading to close attention from the user to the state of the robot. Huggable is in a separate class–it does invite close touch, but also engages the user in a dialog. As a teleoperated robot, it also does not have a clear pre-defined interaction loop. We include it to show that one could incorporate intimate space characteristics in a robot that is also aimed at the personal zone of interaction.

| SAR Name | Image | Persona/Role | Touch affordances | Feedback mechanisms | Interaction loop |
|---|---|---|---|---|---|
| Paro [38, 52] | 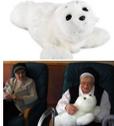 | Baby seal (emotional support animal) | Soft fur, able to respond to touch on body | Mechatronic movement; cooing sounds | SAR responds to light, and sound as well as touch; a range of responses meant to imitate engaging a trusting baby animal |
| Qoobo [18] | 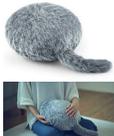 | Cushion with tail (catlike) | Soft fur, able to respond to touch on body | Mechatronic movement (tail swishes at various speeds inviting stroking) | Attract 'tail swish', positive and negative tail movement responses (imitating a cat that likes, then gets overloaded with touch) |
| Haptic Creature [5, 69, 70] | 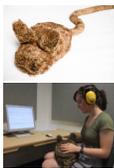 | Research platform, resembles a large rodent | Soft fur, able to respond to touch on back | 'Breathing' motion, variable ear stiffness, purring | Breathing, ears, purring aimed at communicating preferred touch to user |
| Huggable [40] | 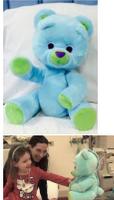 | Bear, for sympathetic discussion and touch | Soft fur, able to respond to hug or hand squeeze | Mechatronic movements; screen-based eyes; voice (tele-operated) | Tele-operated combining touch and conversational interaction |
| Research Prototype | 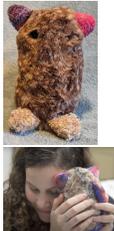 | Research platform, emotional support animal | Soft fur, able to respond to touch on back, ears, feet, tail | Haptic feedback (anxious heartbeat, purring) | Haptic feedback meant to invite user to sooth the creature into calm purring |
| Commercial Design | 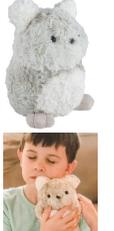 | Commercial design, emotional support animal | Soft fur, able to respond to touch on body and ears | Haptic feedback (anxious heartbeat, purring) and soft noises | Haptic and auditory feedback meant to invite user to sooth the creature into calm purring |

Table 2. Comparison of intimate-space robots.

Overall, one can see from this set of examples, that there seems to be a class of intimate-space robots that can be fruitfully deployed in socially-assistive situations. In the case of Paro, this is to soothe and engage elders who cannot manage a real pet. Qoobo is a sort of novelty version of this, that makes playful reference to the mercurial moods of cats. In the case of our research prototype and final commercial design, these intimate space design characteristics have been deployed to provide a safe opportunity for self-soothing for children, who can take care of the creature and thus also help themselves. The fact that several research and commercial examples have converged upon these common design characteristics suggests that they have merit when designing for intimate-space SARs. We see this as an emergent useful class of robots worthy of further study and development.

## 11 CONCLUSIONS AND FUTURE WORK

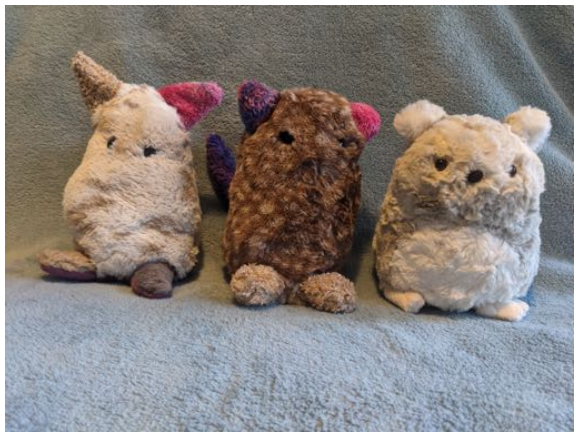

Fig. 19. All three prototypes next to each other from left to right in order of evolution.

We presented a Research-through-design [20] case study focused on creating a child-led, situated intervention to help with emotion regulation, in the form of an intimate-space tangible device best understood as a socially assistive robot. We described the design process and initial evaluation of the device, and the translation of the device into a more robust research product/commercial prototype that allows us to work at the greater scale required of wellbeing interventions (e.g. RCTs). Thanks to the availability of the commercial design, we have been able to engage with multiple partner organizations and populations, including deployments with over 30 mental health clinicians and ethicists, as well as organisations supporting youth who self-harm, providing parenting training to marginalised populations, or those providing support to fostered and looked after children. The research team is currently using the company's current product design to understand the efficacy and mental effects of the underlying psychological interventions. These include an early, medium-scale RCT trial in the US (n=134, Purrble vs active control, primary outcome daily measures of child emotion regulation), as well as a deployment with 50-80 Oxford University students recruited through the university Counselling Service. We are also working with the research toolkit to begin constructing a series of studies that will both enable us to investigate the mechanistic aspects of the intervention (e.g., psychological impacts of alternative haptic patterns and interaction gameplay), gather granular data on the emotion regulation skills development, as well as start examining further expansions of the current intervention model (e.g., by including cognitive training that is associated with the experiential effects of Purrbles). Working together has allowed us to continue research at scale and also, to create an innovative intervention mechanism and a platform that is generating much interest with our clinical partners, while enabling further exploration and blue-sky thinking.

In this paper, we have presented lessons learned from the translation process and recommended some translation strategies. We also introduced a design space of intimate-space socially assistive robots. The research findings are relevant to HRI researchers, as well as those in HCI designing tangibles, and researchers interested in emotion regulation support. The work could also provide valuable guidance to those interested in scaling up work with artifacts and looking to partner (or self-source) translation from research prototypes to research (or commercial) products.

## ACKNOWLEDGMENTS

Thanks to Committee for Children for their generous support of early prototyping and field studies of the device.

We really appreciate the contributions of our undergraduate researchers that helped drive this project.